# Polarization-independent reconfigurable frequency selective rasorber/absorber with low insertion loss

Jing Yuan, Xiangkun Kong, Xuemeng Wang, Shunliu Jiang, Lingqi Kong

A polarization-independent reconfigurable frequency selective rasorber (FSR)/absorber with low insertion loss based on diodes is proposed in this paper. The presented structure consists of a lossy layer based on square loops and a bandpass frequency-selective surface. These two layers are separated by an air layer. Each layer has an embedded bias network that provides the bias voltage to the diodes through metallic via. This configuration can avoid undesirable effects associated with the additional biasing wire. When the diodes are in off-state, the structure is in FSR mode and exhibits a transmission window at 4.28GHz with only 0.69dB insertion loss (IL) within the absorption bands. While diodes are in on-state and the structure switches to absorber mode, it achieves perfect absorption with absorptivity of over 90% ranging from 2.8 to 5.2 GHz. An equivalent circuit model (ECM) is developed to analyse the physical mechanism of the structure. A prototype of the proposed architecture is fabricated and measured, where reasonable agreements between simulations and measurements are observed, verifying the effectiveness of this design.

*Introduction:* Frequency selective surface (FSS) has been widely applied in communication systems and radome designs because of its spatial filtering characteristics [1]. The bandpass FSSs can transmit the signals in the operating band and reflect the out-band signals directly, which may be detected potentially. In order to overcome the problem, the bandpass FSS combining with absorber is designed to absorb out-band signals, which could reduce the Radar Cross Section (RCS) and improve the security of systems. These designs are termed as frequency selective rasorber (FSR) [2].

Recently, many designs of FSR have been proposed, which mainly include three kinds: absorption-transmission (A-T) [3-4], transmission-absorption (T-A) [5] and absorption-transmission-absorption (A-T-A) [6-9]. The typical structure consists of a lossy layer and a lossless layer. The lossy layer possesses an absorption property loaded with lumped resisters. And the lossless layer usually is a bandpass FSS which can provide a transmission window. However, with the fast development of communication systems, the requirement for the multifunctional FSR needs to adapt to different situations. The transmission window of the FSR can be switched by embedding some controllable devices (such as PIN diode, MEMS). The functions of the structure are dependent on the state of diodes, which can change from FSR mode to absorber mode [10-11]. When the structure is in FSR mode, it can transmit the wave at the transmission band locating inside the absorption band; while in the absorption mode, the signals are absorbed in the whole working band. In [10], a switchable FSR/absorber is designed, but it is polarization-sensitive owning to the fact that the complex feeding network often destroys the symmetry of the structure. To overcome this problem, a dual-polarized switchable FSR/absorber using a symmetrical feeding network is presented in [11]. However, the more lumped resistors will result in higher insertion loss

In this letter, a polarization-independent reconfigurable frequency selective rasorber (FSR)/absorber with low insertion loss based on the PIN diode switching technique is proposed. The biggest advantage of this design is that the square loops with metal lines are mutually orthogonal on both sides of lossy layers, which are connected by via and embedded PIN diodes. The slot between square loops in the lossy layer and the square-gap in the lossless layer generate the same transmission band. By embedding PIN diodes into the slot and gap, the structure can switch between FSR mode and absorption mode dependent on the state of diodes. And the structure itself is used as a feeding network which guarantees the whole structure is symmetric and polarization insensitive. In addition, the low IL is attained because the less number of lumped resistors are used on the lossy layer and the most energy of electromagnetic waves can transmit through the structure. The simulated results agree well with the measured results.

*Basic structure design and simulation:* The unit cell of the proposed structure is illustrated in Fig. 1*a*, which consists of a lossy layer, a lossless layer and an air layer between them. Fig.1*b* shows the top view of the lossy layer, which includes two metallic lines and a square loop embedded with lump resistors and PIN diodes. The back structure of the lossy layer in Fig.1*c* is the same as the top structure after being rotated 90°, which guarantees the polarization insensitivity. Moreover, the metallic lines and loops connected by metallic via are also used as a bias network to provide voltage to diodes which avoid the effect of extra bias wires. The lossless layer (Fig.1*d*) is a bandpass FSS loaded with diodes and its bias grids are designed on the back of the lossless layer which is exhibited in Fig.1*e*. The design of the metal patch in the grids is to move the reflection frequency point generated by the via to the high frequency, thereby expanding the absorption bandwidth. The metal and lump elements are printed on an F4B substrate layer with a relative permittivity of 3.5. All the parameters are optimized as follows: $P$=26mm, $t_1$=$t_3$=0.5mm, $t_2$=14.8mm, $L_1$=11mm, $w_1$=$w_2$=$w_3$=0.5mm, $w_r$=0.2mm, $L_2$=17mm, $w_k$=0.2 mm and $w_p$=0.5mm. The PIN diodes (SMP1320-079LF) are chosen from SKYWORKS.

The S-parameters are simulated by commercial full-wave simulation software (CST STUDIO SUITE) with unit cell boundary conditions. Fig. 2*a* exhibits the simulation results when the diodes are in off-state. And the structure works as an FSR (A-T-A). It can be seen that at 4.28 GHz, there is a transmission window with an IL value of 0.69 dB between the higher and lower absorption bands. The two absorption bands can reduce the out-band RCS, which protect the antenna systems more securely. When the diodes are triggered to on–state, the structure switches from FSR mode to absorber mode, as shown in Fig.2*b*. It possesses perfect absorption performance with absorptivity over 90% from 2.8GHz to 5.2GHz. From Fig.2*a* and 2*b*, it is observed that the structure has almost the same performance regardless of the incidence of TE wave or TM wave.

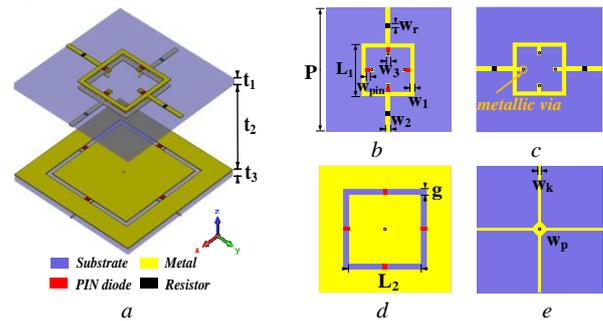

**Fig. 1** *The basic unit cell structure design.*
*a The whole structure of the unit cell.*
*b The top view of the lossy layer.*
*c The back view of the lossy layer.*
*d The top view of the lossless layer.*
*e The back view of the lossless layer.*

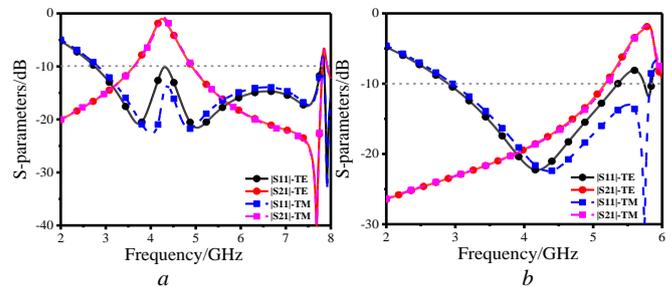

**Fig. 2** *The simulation results of the structure.*
*a When the diodes are in off-state.*
*b When the diodes are in on-state.*

The surface current distributions at 4.28GHz on the top layer and bottom layer are simulated to analyze the principle of work, as shown in Fig.3. In Fig.3a, when diodes are in off-state, the strong surface currents flow along the slot between two square loops on both sides of lossy layer and square-gap on the lossless layer, which can generate the same parallel resonance and offer the transmission window. Meanwhile the currents flow through the loaded resistors when the diodes switch to on-state, the energy can be dissipated by the resistors as shown in Fig.3b.

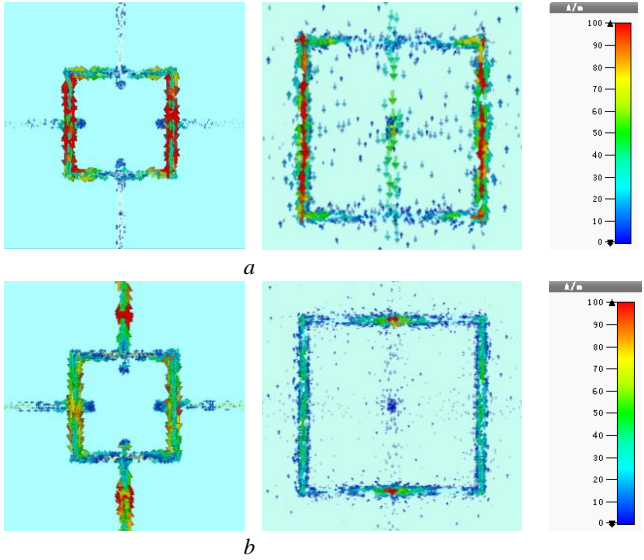

**Fig. 3** *Surface current distributions at 4.28GHz.*
*a The surface current distributions on lossy layer and lossless layer when diodes are in off-state.*
*b The surface current distributions on lossy layer and lossless layer when diodes are in on-state.*

*Equivalent circuit model and analysis*: To fully understand the physical mechanism of the reconfigurable FSR/absorber, a two-port ECM is set up in the ADS software to analyse it as shown in Fig.4*a*. Based on the ECM of the square loop and bandpass FSS in [12], the lossy layer mainly consists of a series–parallel $C_1$, $L_2$ and $L_3$ resonator. The PIN diode is parallel to $C_1$ because it is located between the two loops. $C_g$ represents the coupling of the outer loops on the neighboring two unit cells. A parallel $C_2L_5$ models the bandpass layer. When the diodes are in off-state, the series–parallel resonator on the lossy layer and parallel resistor on lossless layer generate the same transmission band; when the diodes turn to on-state, the bandpass provides metal ground which combined with lossy layer to achieve absorption performance. And the $L_B$ represents the effect of wire grids at the back of the lossless layer. The equivalent inductances and capacitances in the ECM can be calculated from the equivalent components [13-14]. However, the calculated electrical parameters are not very accurate because the coupling between lump elements are not considered in ECM, which need to optimize. The final optimized equivalent electrical parameters are as follows: $L_1$=0.95nH, $C_g$=0.053pF, $R_1$=80Ω, $L_2$=0.2nH, $C_1$=0.31pF, $L_v$=6.1nH, $L_3$=2.62nH, $C_{off}$=0.102pF, $R_{on}$=2.35Ω, $L_{on}$=0.75nH, $R_2$=420Ω, $L_4$=5.1nH, $C_2$=0.715pF, $L_5$=2.33nH, $L_B$=7.2nH, $Z_{sub}$=$Z_0/\sqrt{\varepsilon_r}$. $Z_0$ and $Z_{sub}$ are the characteristic impedance of free space and dielectric spacer, respectively. $\varepsilon_r$ is the relative permittivity of the substrate.

Fig.4*b* and Fig.4*c* show the scattering parameters of the proposed structure from CST and ADS under normal incidence. The simulated results of CST agree well with that of ECM calculated by ADS. It can be seen that when the diodes are in the off-state, the structure performs as an FSR with a transmission window inside the absorption band, and the insertion loss of the transmission window is relatively low. And when the diodes are in the on-state, the structure changes from FSR to absorber that can realize perfect absorption at the whole working band.

*Experimental verification:* To demonstrate the performance of the proposed reconfigurable FSR/absorber, a prototype is fabricated and tested in a specially designed TEM cell. The TEM cell is an open-ended waveguide structure which can support quasi-TEM mode propagating along z-direction and can offer an efficient way to characterize periodic structures along x-direction according to the principle of mirror image. As shown in Fig. 5*a*, the experiment system is set up in the anechoic chamber. A vector network analyser (VNA) is used to transmit and receive electromagnetic waves. And the time gating is employed to reduce the multiple reflection effects. Fig.5*b* shows the photograph of the fabricated prototype, which consists of 1×9 unit cells with a size of 26×234 mm$^2$. The metallic structures are printed on 0.5mm-thick F4B using printed circuit board (PCB) technology. A 14.8 mm-thick foam spacer and nylon screws are employed to separate the two functional layers and fix the whole structure. The DC regulated power supply is used to excite the diodes to change functions of the structure.

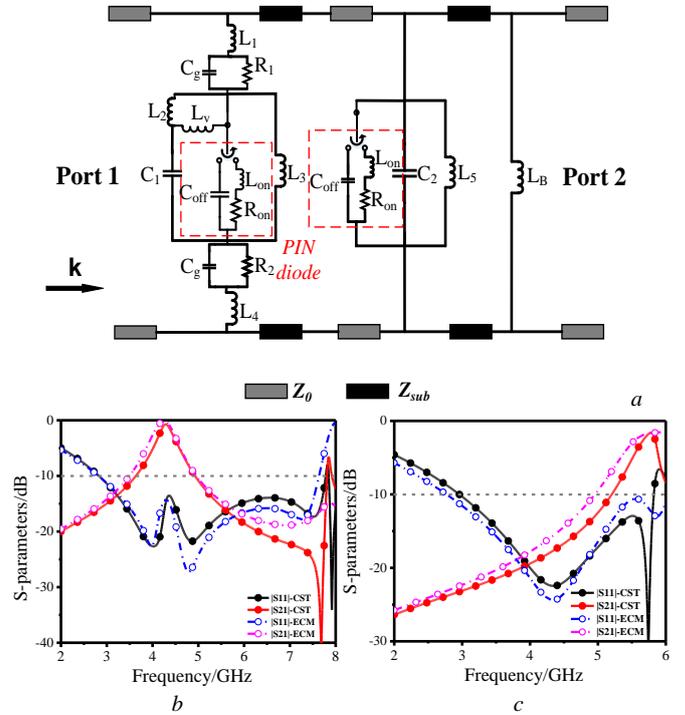

**Fig. 4** *The ECM of the structure and simulated results.*
*a The ECM of the whole structure.*
*b The compared simulation results when diodes are in off-state.*
*c The compared simulation results when diodes are in on-state.*

The measured results and simulated results are plotted in Fig.6. As illustrated in Fig.6*a*, when PIN diodes are in off-state, the frequency deviation generates and the measured IL at 4.32GHz is about 1.2dB which is slightly higher than that of simulation. And when the PIN diodes are in on-state, the whole structure has more than 90% absorptivity from 2.6GHz to 5.3GHz as shown in Fig.6*b*. There are some differences between simulation results and measurement results. The relevant lumped elements and diodes are equivalent to ideal circuit models in the simulation according to the datasheet, which ignores the parasitic effects of actual encapsulation. Intrinsic resistance of diodes, inaccurate sample fabrication, value variations also lead to these deviations.

Table 1 demonstrates the performance comparison with existing designs. It is clearly observed that this work design has many advantages in terms of polarization independence, low IL, less number of lumped resistors, reconfigurable functionality and feeding network integration. Moreover, compared with other reconfigurable designs, the proposed structure possesses the outstanding characteristics of polarization independence and much lower IL at transmission window. But it is also noted that FBW of the absorber mode is slightly narrower than the designs of [10] and [11]. How to expand the bandwidth may be a meaningful subject for a future study.

*Conclusion*: In this letter, a polarization-independent reconfigurable frequency selective rasorber/absorber with low insertion loss is presented. The overall structure mainly consists of a top lossy layer and a bottom lossless bandpass FSS, where the PIN diodes mounted across the slot and gap. An integrated bias feeding configuration is designed to control the diode arrays without negative effects on the performance of the whole structure, which guarantees the symmetry of the structure and

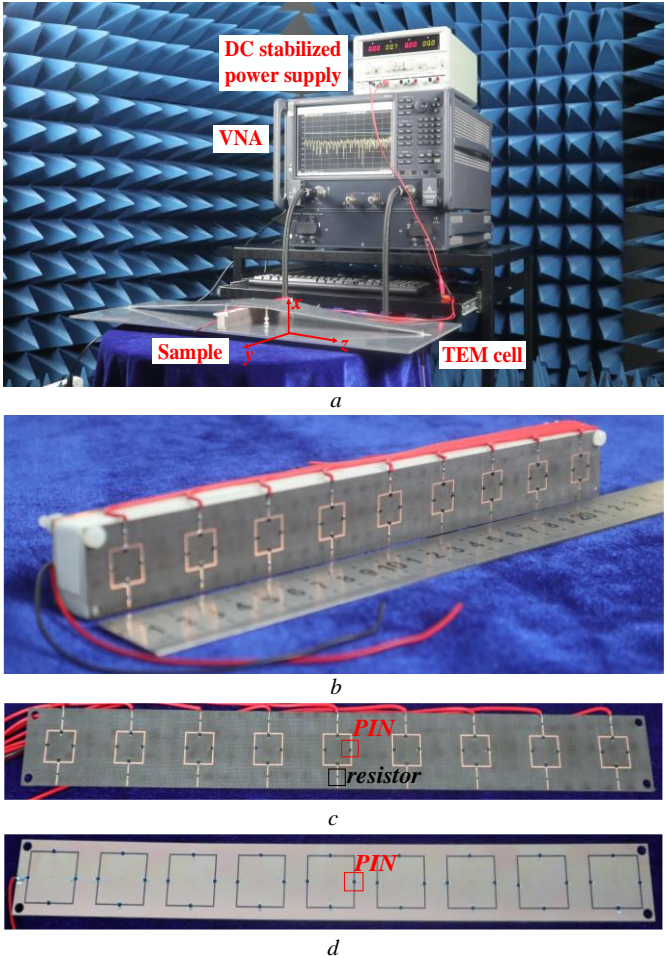

**Fig. 5** *Experiment system and prototype of the structure*
*a Experiment system configuration of the measurement.*
*b The view of the fabricated sample.*
*c The view of the lossy layer.*
*d The view of the lossless layer.*

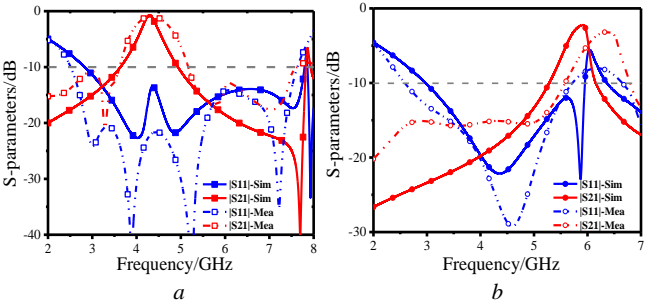

**Fig. 6** *The results comparison between simulation and measurement*
*a Measured and simulated results when the diodes are in off-state.*
*b Measured and simulated results when the diodes are in on-state*

**Table 1** Performance Comparison

| Ref. | IL$_{min}$ (dB) | NOR | FBW(%) | Pol. | Rec. |
|---|---|---|---|---|---|
| [6] | 0.34 | 6 | FSR(92.6); --- | Dual | NO |
| [7] | 0.7 | 8 | FSR(101); --- | Dual | NO |
| [8] | 0.21 | 4 | FSR(94.5); --- | Dual | NO |
| [9] | 0.2 | 8 | FSR(108); --- | Dual | NO |
| [10] | 0.6 | 3 | FSR(106.2); Ab. (99.6) | Single | YES |
| [11] | 1.7 | 8 | FSR(124); Ab. (121) | Dual | YES |
| **This work** | **0.69** | **4** | **FSR(97.2 ); Ab. (60)** | **Dual** | **YES** |

IL$_{min}$ is the minimum insertion loss at the frequency of transmission band, FBW represents fractional bandwidth of −10 dB reflection. NOR=Number of resistors, Ab.=Absorber, Pol.= Polarization, Rec.= Reconfigurable.

polarization insensitivity. Reconfigurable functionality between FSR mode and absorber mode is realized by controlling the state of diodes. The IL of transmission window is less than 1dB and a prototype is fabricated to verify the effectiveness of the design. In a complex electromagnetic environment, this design has potential application prospects in stealth technology, communication systems, electromagnetic compatibility and specific radome design.

*Acknowledgments:* This work was supported by the Fundamental Research Funds for the Central Universities (No.kfjj20190406), the Postgraduate Research & Practice Innovation Program of Jiangsu Province under Grant (SJCX20_0070), National Natural Science Foundation of China (61471368) and by Open Research Program in China's State Key Laboratory of Millimeter Waves (Grant No. K202027).

Jing Yuan, Xiangkun Kong, Xuemeng Wang, Shunliu Jiang, Lingqi Kong (*Key Laboratory of Radar Imaging and Microwave Photonics, Nanjing University of Aeronautics and Astronautics, Nanjing 210016, China*).
Xiangkun Kong (*The State Key Laboratory of Millimeter Waves, Southeast University, Nanjing 210096, China*).

E-mail: xkkong@nuaa.edu.cn